\documentclass[10pt, conference]{IEEEtran}
\usepackage{amsfonts}
\usepackage{amssymb}
\usepackage{enumerate}
\usepackage{color,xcolor}
\usepackage{graphicx}
\usepackage{multirow,tabularx}
\usepackage{subfigure}
\usepackage{hyperref}

\begin{document}

\title{A Low-Complexity Joint Detection-Decoding Algorithm for Nonbinary LDPC-Coded Modulation Systems}

\author{
\authorblockN{Xuepeng~Wang,~Baoming~Bai}
\authorblockA{State Key Lab. of ISN, Xidian University\\
Xi'an 710071, China\\
E-mail: wangxuepengxd@gmail.com\\
~~bmbai@mail.xidian.edu.cn} \and
\authorblockN{Xiao~Ma}
\authorblockA{Department of ECE, Sun Yat-sen University\\
Guangzhou, GD 510275, China\\
E-mail: maxiao@mail.sysu.edu.cn} }


%
\maketitle

\begin{abstract}
In this paper, we present a low-complexity joint
detection-decoding algorithm for nonbinary LDPC coded-modulation
systems. The algorithm combines hard-decision decoding using the
message-passing strategy with the signal detector in an iterative
manner. It requires low computational complexity, offers good
system performance and has a fast rate of decoding convergence.
Compared to the $q$-ary sum-product algorithm (QSPA), it provides
an attractive candidate for practical applications of $q$-ary LDPC
codes.
\end{abstract}


%
\IEEEpeerreviewmaketitle

\section{Introduction}
Nonbinary low-density parity-check~(LDPC) codes were first
introduced by Gallager in~\cite{Gallager1963} based on modulo
arithmetic. In~\cite{Davey1998}, Davey and MacKay presented a class
of nonbinary LDPC codes defined over finite field GF($q$) with
$q>2$. They also introduced a sum-product algorithm (SPA) for
decoding $q$-ary LDPC codes, named QSPA. Now, it has been shown
that nonbinary LDPC codes have better performance than binary LDPC
codes~\cite{Davey1998},~\cite{Hu2004}, especially when combined
with higher-order modulations. Recently, a surge appears in
the study of nonbinary LDPC codes~\cite{Declercq2007,Song2006ISIT,Rathi2005,Bennatan2006,Li2009}.

However, the advantages of nonbinary LDPC codes over its binary
counterpart are balanced by their higher decoding complexity. To
reduce the decoding complexity, Davey and MacKay proposed a more
efficient QSPA, called fast Fourier transform based QSPA
(FFT-QSPA), for decoding LDPC codes over
GF($2^p$)~\cite{MacKay2000}. Moreover,
a simplified decoding algorithm called extended min-sum (EMS) was
proposed by Declercq and Fossorier in~\cite{Declercq2007} to further reduce decoding
complexity. It provides a good candidate for decoding $q$-ary LDPC
codes with small $q$. For larger field size (say, $q>32$), the sort
operations required by the EMS algorithm will incur higher
complexity. As a result, the decoding complexity is still a
concern for practical implementation of $q$-ary LDPC coded
systems.

Most recently, Mobini \textit{et al}~\cite{Mobini2009} and Huang
\textit{et al}~\cite{Huang2009} developed reliability-based
decoding algorithms for binary LDPC codes with low complexity.
Motivated by their work and ~\cite{Nouh2002}, this paper will
explore hard-decision based decoding for $q$-ary LDPC codes, and
present a low-complexity joint detection-decoding algorithm for
$q$-ary LDPC-coded modulation systems, which provides efficient
trade-off between system performance and implementation complexity.

The algorithm is devised to combine the simplicity of
hard-decision decoding with the good performance of
message-passing algorithms. In the proposed scheme, signal
detection and decoding are integrated as a whole, and the input
signal vector to detector is updated in an iterative way. At each
iteration, the updated hard-decision results from detector are
delivered to the LDPC decoder which performs hard-decision
decoding using message-passing algorithm. The output of decoder is
then fed back to detector, with which the received signal points
are updated such that they are progressively close to the
transmitted signals in observation space. This can be viewed as an
iterative denoising processing. Compared to the FFT-QSPA, the
proposed algorithm requires lower computational complexity and has
fast rate of decoding convergence.


\section{Nonbinary LDPC-Coded Modulations}

\subsection{System Model}
\begin{figure}
    \center
    \includegraphics[width=7cm]{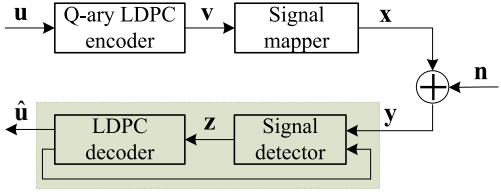}
    \caption{A nonbinary LDPC-coded modulation system} \label{LDPCCM}
\end{figure}

The nonbinary LDPC-coded modulation system under consideration is
shown in Fig.~\ref{LDPCCM}. Assume that an LDPC code
$\mathcal{C}[N,K]$ over GF($q$) with $q>2$ is used in conjunction
with a two-dimensional signal constellation $\mathbf{\mathcal{X}}$
of size $|\mathbf{\mathcal{X}}|$. The input vector of information
symbols, $\mathbf{u}\in \mathrm{GF}(q)^K$, is first encoded by the
LDPC encoder into a codeword $\mathbf{v}=(v_0,v_1,...,v_{N-1})\in
\mathcal{C}$. The corresponding code rate $R_c=K/N$. The codeword
$\mathbf{v}$ is then mapped to $\mathbf{\mathcal{X}}$, producing
the modulated signal vector $\mathbf{x}=(x_0,x_1,...,x_{N-1})$
with ${x_j}=\mathcal{M}(v_j)\in \mathcal{X}$, where
$\mathcal{M}(\cdot)$ stands for the signal mapping function. In
this paper, we always assume the constellation size is equal to
the finite field size, i.e., $|\mathbf{\mathcal{X}}|=q$. The
spectral efficiency for this coded-modulation system is
    \begin{equation}
        \rho= R_c {\mathrm{log}_{2}}|\mathbf{\mathcal{X}}|~~~~\mathrm{bits/signal}.
    \end{equation}
Suppose that the complex signal vector $\mathbf{x}$ is transmitted
over the AWGN channel. The received vector
$\mathbf{y}=(y_0,y_1,...,y_{N-1})$ is then given by
    \begin{equation}
        y_j=x_j+n_j,~~j=0,1,...,N-1,
    \end{equation}
where ${n_j}\sim \mathcal{CN}(0,N_0)$ are independent and
identically distributed complex Gaussian random variables with
zero mean and variance $N_0/2$ per dimension. Denote by
$E_s=\mathbf{E}[{|x_j|}^{2}]$ the average energy per transmitted
signal. Then the average received signal-to-noise ratio (SNR) is
\begin{displaymath}
\mathrm{SNR}={E_s}/{N_0}=\rho {E_b}/{N_0},
\end{displaymath}
where $E_b$ denotes the average energy per information bit.

In this paper, we consider a hard-decision based iterative
detection-decoding strategy. In each iteration the signal detector
makes hard-decision about $\mathbf{v}$ based on the updated
received vector $\mathbf{y}$, producing vector
$\mathbf{z}=(z_0,z_1,...,z_{N-1})$ with $z_j\in$ GF($q$); then the
decoder performs hard-decision decoding with $\mathbf{z}$ as the
input. The hard extrinsic-information produced by the decoder is
then fed back to the signal detector to update $\mathbf{y}$. We
will show that with this decoding strategy, good system
performance can be achieved with reduced complexity.

\subsection{LDPC Codes over GF($q$)}
A $q$-ary LDPC code $\mathcal{C}$ of length $N$ over GF($q$) is
given by the null space of a sparse $M\times N$ parity-check
matrix $\mathbf{H}=[h_{i,j}]$ over GF($q$), where $M$ is the
number of check equations and $M=N-K$ if $\mathbf{H}$ is full
rank. Let $\mathbf{v}=(v_0,v_1,...,v_{N-1})$ be a codeword in
$\mathcal{C}$. Then the parity-check constraints can be expressed
as $\mathbf{vH}^T=\mathbf{0}$, or
    \begin{equation} \label{j}
        \sum_{j=0}^{N-1}\ h_{i,j}v_{j}=0,~~i=0,1,...M-1,
    \end{equation}
where the operations of multiplication and addition are all
defined over GF($q$). If the matrix $\mathbf{H}$ has constant row
weight $d_c$ and constant column weight $d_v$, then the
corresponding code is called a $(d_v,d_c)$-regular $q$-ary LDPC
code.

    \begin{figure}\center
        \includegraphics[width=9cm]{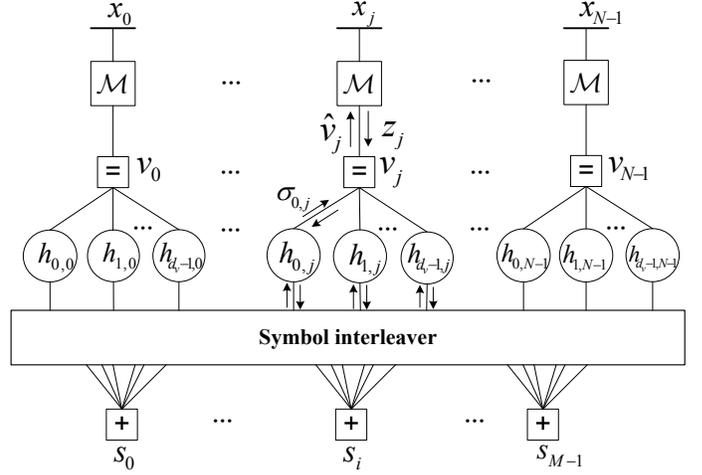}\\
        \caption{Forney-style factor graph of the nonbinary LDPC-coded system}\label{FGgraph}
    \end{figure}

Similar to its binary counterpart, a $q$-ary LDPC code can also be
described using a Forney-style factor graph~\cite{Forney01},
as depicted in Fig.~\ref{FGgraph}, where $\mathcal{M}$ denotes the mapper/demapper.
The graph has $N$ variable nodes corresponding to coded symbols,
and $M$ check nodes
corresponding to parity-check equations. For convenience, we
define the following two index sets
    \begin{equation}\label{Nvj}
        {\mathcal{N}_v}(j)=\{i|h_{i,j}\neq 0,~0\leq i<M\},
    \end{equation}
    \begin{equation}\label{Nci}
        {\mathcal{N}_c}(i)=\{j|h_{i,j}\neq 0,~0\leq j<N\}.
    \end{equation}

\section{Joint Detection-Decoding Algorithm for $q$-ary LDPC-Coded Systems}

In this section, we will develop an iterative joint
detection-decoding algorithm for the LDPC-coded modulation system
shown in Fig.~\ref{LDPCCM}. The whole algorithm is a hard-decision
based message-passing algorithm (MPA) operating on the factor
graph. Assume that a $(d_v,d_c)$-regular $q$-ary LDPC code is
used.

\subsection{Signal Detection and Message-Passing Based on Hard Information}
\label{sec:3.1} Let
$\hat{\mathbf{x}}=(\hat{x}_0,\hat{x}_1,...,\hat{x}_{N-1})$ with
$\hat{x}_j \in \mathcal{X}$ denote the estimate of $\mathbf{x}$
made by the signal detector based on the received vector
$\mathbf{y}$. Assume that all the constellation points in
$\mathcal{X}$ are used equal likely. Then with the maximum
likelihood decision rule, the detected signal $\hat{\mathbf{x}}$
is given by
    \begin{equation}\label{detection}
        \hat{x}_j=\mathrm{arg min}_{x\in \mathcal{X}}\|y_j-x\|, ~~j=0,1,...,N-1,
    \end{equation}
where $\|\cdot\|$ denotes the Euclidean ($l_2$) norm. The input of
the decoder $\mathbf{z}$ is simply the
demapping output of $\hat{\mathbf{x}}$, i.e.,
    \begin{equation}\label{demapping}
        z_j=\mathcal{M}^{-1}(\hat{x}_j)\in \mathrm{GF}(q),~~j=0,1,...,N-1.
    \end{equation}
The vector $\mathbf{z}$ is a codeword if and only if its $M$-tuple
syndrome $\mathbf{s}=(s_0,s_1,...,s_{M-1})$ equals $\mathbf{0}$,
i.e.,
    \begin{equation}\
        \mathbf{s}\equiv\mathbf{z}\mathbf{H}^T=\mathbf{0}.
    \end{equation}
For $0\leq i < M$, the component $s_i$ of $\mathbf{s}$ is given by
    \begin{equation}\label{syndrome}
        s_i=\sum_{j\in{\mathcal{N}_{c}}(i)}\ h_{i,j}z_{j},
    \end{equation}
which is called a \emph{check-sum} of received symbols. A received
symbol $z_j$ is said to be checked by $s_i$ if $h_{i,j}\in
\mathrm{GF}(q)$$\setminus\{0\}$. From~(\ref{syndrome}) with
$s_i=0$, the estimate of $v_j$ given by other variable nodes
participating in the check-sum $s_i$ can be expressed as
    \begin{equation}\label{extrinsic}
        \sigma_{i,j}=-h_{i,j}^{-1}(\sum_{j'\in{\mathcal{N}_{c}}(i)\setminus j}\ {h_{i,j'}} z_{j'}).
    \end{equation}
This is the \emph{update rule for check-to-variable node message}.
Since the column weight of $\mathbf{H}$ is $d_v$, every symbol
$v_j$ can receive $d_v$ estimates along the set of edges
$\{(i,j)|i\in {\mathcal{N}_v}(j)\}$, as depicted in
Fig.~\ref{FGgraph}. As can be seen from~(\ref{syndrome})
and~(\ref{extrinsic}) , only hard information propagates between
variable and check nodes, i.e., the variable nodes send their
hard-decision decoded symbols to the check nodes, and the check
nodes simply compute the syndromes and send estimates back to
their adjacent variable nodes. The estimate $\sigma_{i,j}$ for
$v_j$ can be considered as the extrinsic
information~\cite{Huang2009}.

\subsection{Iterative Detection-Decoding}
\label{sec:3.2}
We now proceed to consider the update rule for variable nodes.
Refer to Fig.~\ref{FGgraph}. Assume that a variable node has received
the extrinsic information from adjacent check nodes. Different
from general message-passing algorithms, we will use this
information to update the received samples to improve the
reliability measure of received signal. To do this, an iterative
process is performed between variable nodes and signal detector.
For simplicity, the proposed iterative joint detection-decoding
algorithm will be referred to as IJDD hereafter.

In the following, we first introduce some notations used for the
IJDD algorithm. Let $k_{max}$ be the maximum number of iterations
to be performed. For $0 \leq k < k_{max}$, let:
\begin{itemize}
    \item $\mathbf{y}^{(k)}=(y_{0}^{(k)},y_{1}^{(k)},...,y_{N-1}^{(k)})$
            be the input vector to the signal detector in the $k$th iteration; $\hat{\mathbf{x}}^{(k)}$ and
            $\mathbf{z}^{(k)}$ be the corresponding detected signal vector and
            the output decision vector.
    \item $\mathbf{s}^{(k)}=(s_{0}^{(k)},s_{1}^{(k)},...,s_{M-1}^{(k)})$
            be the syndrome of $\mathbf{z}^{(k)}$.
    \item $\sigma_{i,j}^{(k)}$ be the extrinsic information for $v_j$
            given by the $i$th check-sum involving $v_j$ in the $k$th iteration.
    \item $\mathcal{D}(y_{j}^{(k)},r)$ denote a valid search sphere of
            radius $r$ centered at $y_{j}^{(k)}$.
    \item ${\vec{\mathbf{L}}_{j}}^{(k)}(\mathbf{p},\mathbf{q})=\mathbf{q}-\mathbf{p}$
            be a correction vector directed from the point $\mathbf{p}$ to the point $\mathbf{q}$.
            For brevity, we will use ${\vec{\mathbf{L}}_{j}}^{(k)}$
            for ${\vec{\mathbf{L}}_{j}}^{(k)}(\mathbf{p},\mathbf{q})$.
    \item $f_j^{(k)}(a)$, $a\in \mathrm{GF}(q)$ denote the number of occurrences of the element $a$ in
            $\{\sigma_{i,j}^{(k)}\}_{i\in \mathcal{N}_v(j)}$.
\end{itemize}
Clearly, $0\leq f_j^{(k)}(a)\leq d_v$ and $\sum_{a\in{\mathrm{GF}(q)}}f_j^{(k)}(a)=d_v$.
$f_j^{(k)}(a)$ indicates a reliability measure for
decoding $z_j$ into the symbol $a$. Let
\begin{equation}
\label{vote}
    a_{max}=\arg\max_{a\in \mathrm{GF}(q)}\{f_j^{(k)}(a)\},
\end{equation}
and
\begin{displaymath}
\Delta f_j^{(k)}=f_j^{(k)}(a_{max})-\max_{a\in \mathrm{GF}(q)\setminus \{a_{max}\}} \{f_j^{(k)}(a)\},
\end{displaymath}
where $a_{max}$ is the element in GF($q$) that has the highest
reliability for $v_j$, and $\Delta f_j^{(k)}$ represents the
difference in number of votes between the two highest-voted
candidates for $v_j$ in the $k$th iteration. With the plurality
voting rule, we choose
\begin{displaymath}
\hat{v}_j^{(k)}=a_{max}.
\end{displaymath}

With the above discussions, the \emph{message update rule for variable nodes} can be formulated as
follows.

\underline{\emph{Message Update Rule for Variable Nodes:}} Make
estimation on $\hat{v}_j^{(k)}$ using~(\ref{vote}) based on
            ${\{\sigma_{i,j}^{(k)}\}}_{i\in {\mathcal{N}_{v}(j)}}$, $j=0,1,...,N-1$,
            and evaluate $f_j^{(k)}(a_{max})$ and $\Delta f_j^{(k)}$.
Then the variable nodes send the triples $(\hat{v}_j^{(k)},\Delta f_j^{(k)},f_j^{(k)}(a_{max}))$,
$0\leq j\leq N-1$, to detector, where
the following operations are performed on $y_j^{(k)}$:
\begin{equation}\label{correction}
y_j^{(k+1)}=y_j^{(k)}+\xi_j^{(k)}{\vec{\mathbf{L}}_j^{(k)}},
\end{equation}
Here ${\vec{\mathbf{L}}_j^{(k)}}$ and $\xi_j^{(k)}$ are given based on $\hat{x}_j^{(k)}$ and
$(\hat{v}_j^{(k)},\Delta f_j^{(k)},f_j^{(k)}(a_{max}))$, $0\leq j\leq N-1$,
which can be described as follows.

\begin{itemize}
\item If $\mathcal{M}(\hat{v}_j^{(k)})\in
\mathcal{D}(y_{j}^{(k)},r)$, then
\begin{equation}
\label{eq-move}
\vec{\mathbf{L}}_j^{(k)}=
\left\{
  \begin{array}{ll}
    \hat{x}_j^{(k)}-y_j^{(k)}, & \mbox{\textrm{ if }$\mathcal{M}(\hat{v}_j^{(k)})=\hat{x}_j^{(k)}$} \\
    \mathcal{M}(\hat{v}_j^{(k)})-\hat{x}_j^{(k)}, & \mbox{\textrm{ if }$\mathcal{M}(\hat{v}_j^{(k)})\neq\hat{x}_j^{(k)}$}
  \end{array}
\right.
\end{equation}

and

\begin{equation}
\label{fraction}
\xi_j^{(k)}=
\left\{
    \begin{array}{ll}
        f_j^{(k)}(a_{max})/{d_v}, &\mbox{\textrm{ if }$\Delta f_j^{(k)}\geq T$} \\
        \Delta f_j^{(k)}/{d_v}, &\mbox{\textrm{ if }$\Delta f_j^{(k)}<T$}
    \end{array}
\right.
\end{equation}

\item Otherwise set ${\vec{\mathbf{L}}_j^{(k)}}=\mathbf{0}$ and $\xi_j^{(k)}=0$.
\end{itemize}

Note that $\mathcal{D}(y_{j}^{(k)},r)$ specifies a region
where $x_{j}$ is located most likely. Only
$\mathcal{M}(\hat{v}_j^{(k)})$ within this region are used to
update $y_{j}^{(k)}$. In this paper we set $r$ to
be 1.415$d_{min}$ and $T$ to be 3, where $d_{min}$ is the minimum Euclidean distance
among constellation points. Moreover, the case of $\mathcal{M}(\hat{v}_j^{(k)})$=$\hat{x}_j^{(k)}$
in~(\ref{eq-move}) indicates that, $y_j^{(k)}$ may be decoded into
$\hat{x}_j^{(k)}$ with high confidence. In other words, $y_j^{(k)}$
is considered as the noisy version of $\hat{x}_j^{(k)}$. However, instead of
instantly setting $y_j^{(k)}$ to be $\hat{x}_j^{(k)}$, cautious shift is operated on
$y_j^{(k)}$ towards $\hat{x}_j^{(k)}$.
While in the case of $\mathcal{M}(\hat{v}_j^{(k)})\neq\hat{x}_j^{(k)}$, the received
signal will be shifted towards the decision boundary of the two candidates,
achieving a trade-off between the two choices.
Here the decision boundary of $\mathcal{M}(\hat{v}_j^{(k)})$
and $\hat{x}_j^{(k)}$ is the bisector perpendicular to $\vec{\mathbf{L}}_j^{(k)}=\mathcal{M}(\hat{v}_j^{(k)})-\hat{x}_j^{(k)}$.

\begin{figure}
\centering
  \includegraphics[width=9cm]{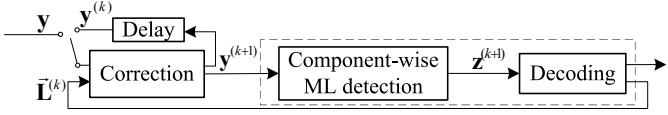}\\
  \caption{Iterative joint detection-decoder.}\label{JDD}
\end{figure}

Then using ${\vec{\mathbf{L}}_j^{(k)}}$ and $\xi_j^{(k)}$,
the detector updates currently input vector according to~(\ref{correction}). Based on
$y_{j}^{(k+1)}$, new hard-decision is made, and the results are
delivered to the variable nodes as the updated message passed from
detector to the decoder. The whole working process of the IJDD algorithm is shown
in Fig.~\ref{JDD}.
%

In summary, the IJDD algorithm can be formulated as follows.

\noindent\rule[0.25\baselineskip]{0.485\textwidth}{0.6pt}

\begin{center}
\textbf{IJDD Algorithm}
\end{center}

\noindent\rule[0.25\baselineskip]{0.485\textwidth}{0.6pt}

\begin{itemize}
    \item \textbf{Initialization.}\\
            Set $k=0$, and ${\mathbf{y}}^{(0)}=\mathbf{y}$.

    \item \textbf{Repeat the following while} $k<k_{max}$
    \begin{enumerate}
        \item [1)] \textbf{Signal detection:}\\
                - For $j=0,1,...,N-1$, make ML decisions based on $y_j^{(k)}$, yielding $\hat{x}_j^{(k)}$ and $z_j^{(k)}$ as in~(\ref{detection}) and
                (\ref{demapping}).\\
                - Pass the message ${\mathbf{z}}^{(k)}$ to LDPC decoder.
        \item [2)] \textbf{Compute syndrome and do check-node update:}\\
                - Compute syndrome vector
                $\mathbf{s}^{(k)}=\mathbf{z}^{(k)}\mathbf{H}^T$.\\
                - If $\mathbf{s}^{(k)}=\mathbf{0}$, then terminate iteration and output $\hat{\mathbf{v}}^{(k)}=\mathbf{z}^{(k)}$ as the decoded
                codeword;\\
                - else for $i=0$ to $M-1$ and $j=0$ to $N-1$,\\
                  compute check-to-variable messages $\sigma_{i,j}^{(k)}$ as
                  in~(\ref{extrinsic}).

        \item [3)] \textbf{Variable-node update and correction:}\\
                For $j=0$ to $N-1$,\\
                - evaluate $(\hat{v}_j^{(k)},\Delta f_j^{(k)},f_j^{(k)}(a_{max}))$ based on $\{\sigma_{i,j}^{(k)}\}$ at variable nodes;\\
                - send the message triples $(\hat{v}_j^{(k)},\Delta f_j^{(k)},f_j^{(k)}(a_{max}))$ to detector and perform signal corrections
                as done in Section~\ref{sec:3.2}.

        \item [4)] $k=k+1$, \textbf{entering next iteration.}

    \end{enumerate}

    \item If $k=k_{max}$, declare a decoding failure.

\end{itemize}

\noindent\rule[0.25\baselineskip]{0.485\textwidth}{0.6pt}

It is worth mentioning that in the IJDD algorithm differs from
QSPA/FFT-QSPA in two aspects: 1) In the IJDD algorithm, only
simple operations such as additions, comparisons, look-up tables,
negligible amount of real operations and finite field
operations are required; 2) The iterative process based on hard
information is performed among detector/demapper, variable nodes
and check nodes, while in the QSPA/FFT-QSPA, soft information
propagates only between variable and check nodes.

From the above, it can be seen that the IJDD algorithm is easy to implement, and
can achieve high speed. However, like existing reliability-based
decoding algorithms, to ensure the reliability of majority voting,
the column weights of $\mathbf{H}$ have to be relatively large. It
seems not easy for randomly constructed $q$-ary LDPC codes to
fulfill this requirement, thus application of IJDD algorithm is
restricted to $q$-ary LDPC codes constructed based on finite
fields~\cite{Zeng2008A} or finite geometry~\cite{Zeng2008B}.

\section{Simulation Results}

In this section, two examples of $q$-ary LDPC codes are provided to
demonstrate the effectiveness of the proposed IJDD algorithm.

\begin{figure}
    \centering
    \subfigure[Received signal space]{
        \label{scatterplot:subfig:a}
        \includegraphics[width=4.5cm]{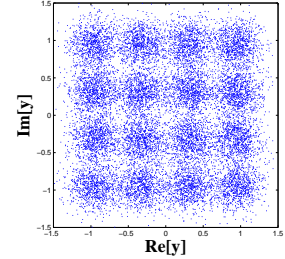}}
    \hspace{1.0cm}
    \subfigure[Corrected signal space]{
            \label{scatterplot:subfig:b}
            \includegraphics[width=4.4cm]{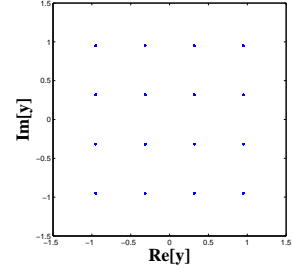}}
  \caption{Scatter plot of signal space before and after decoding with 10 iterations}\label{scatterplot}
  \label{scatterplot:subfig}
\end{figure}

\emph{Example 1:}  Consider a 16-ary (255, 175) regular LDPC code
constructed based on finite fields. The factor graph of this code
has 255 variable nodes and 255 check nodes (including 175
redundant check equations). Both the row and column weights are
16. With the use of 16-QAM signaling over the AWGN channel,
scatter plots for signal vectors before and after decoding using IJDD
with 10
iterations at $E_b/N_0=8.0$ dB are shown in Fig.~\ref{scatterplot:subfig:a} and
Fig.~\ref{scatterplot:subfig:b}, respectively. Clearly, the scatter plot in Fig.~\ref{scatterplot:subfig:b}
corresponds to an ideal signal constellation, i.e., all received
samples have been shifted to the probable originally transmitted
constellation points.

\begin{figure}\center
  \includegraphics[width=7cm]{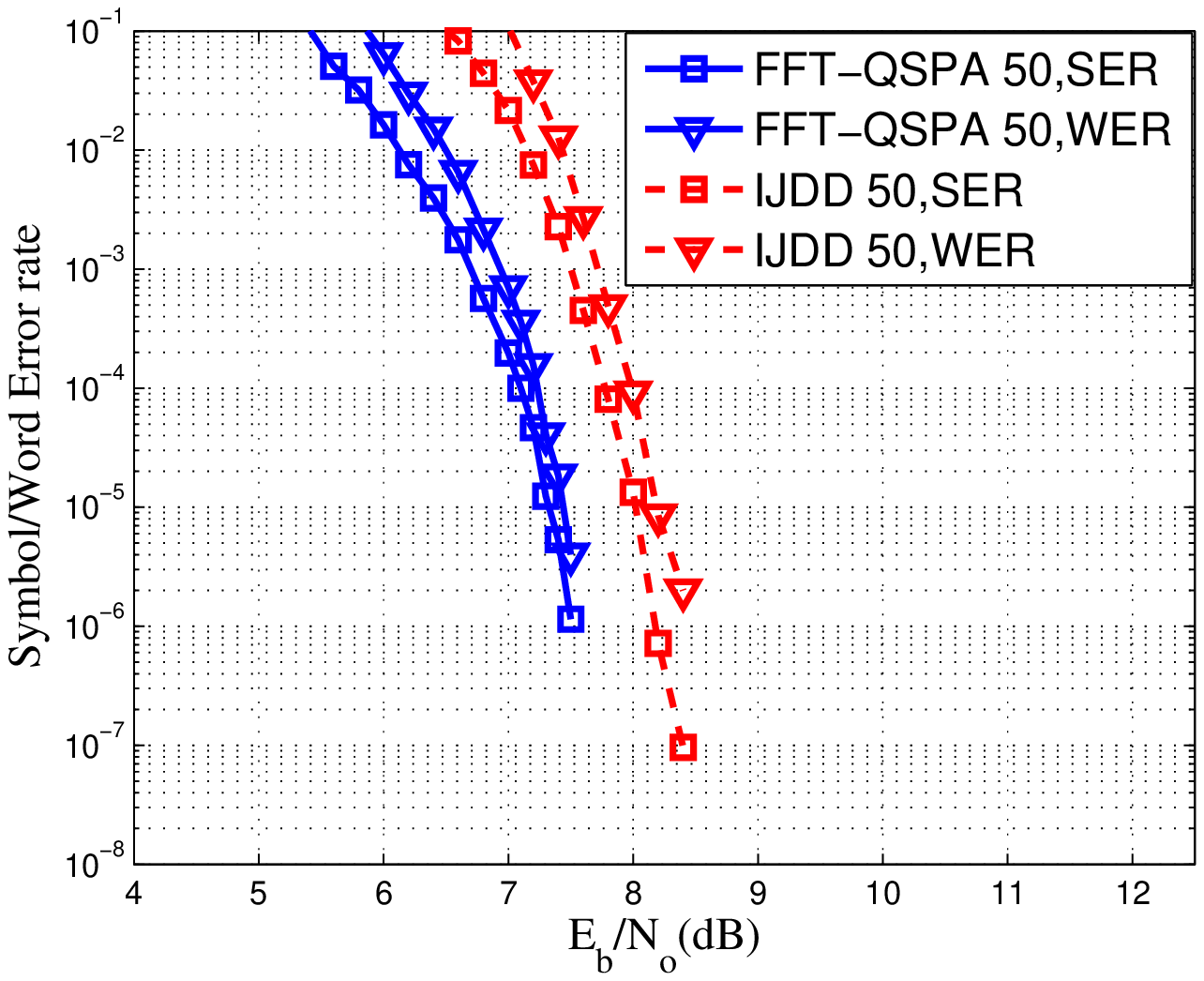}\\
  \caption{Error performance of the 16-ary (255,175) LDPC code decoded with the FFT-QSPA and the IJDD algorithm (16-QAM)}\label{SER_WER}
  \includegraphics[width=7cm]{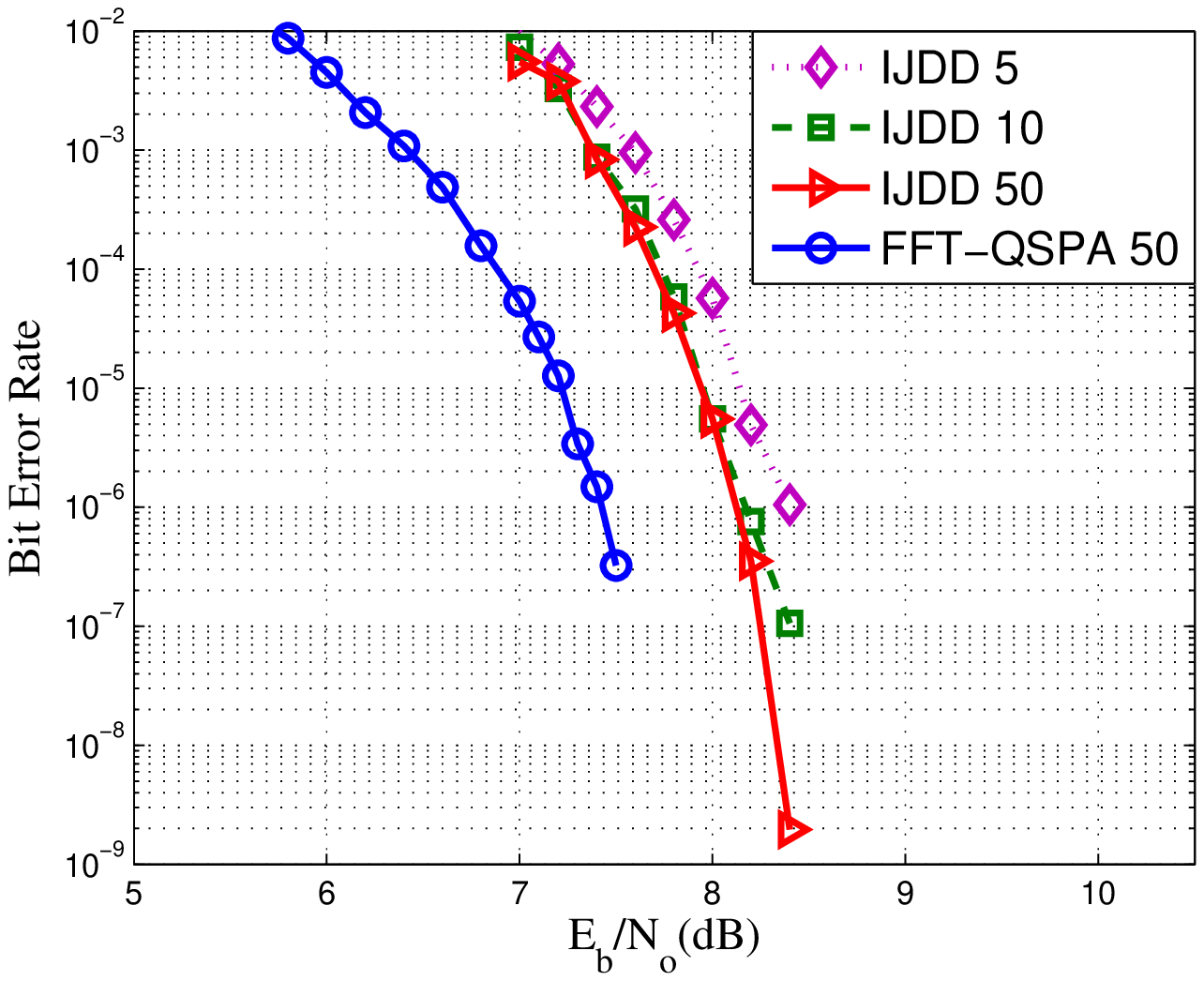}\\
  \caption{Rate of decoding convergence of the IJDD algorithm for decoding the 16-ary (255,175) LDPC code (16-QAM)}\label{Rate_conv}
\end{figure}

Shown in Fig.~\ref{SER_WER} are the symbol and word error performances of this
coded system decoded using the IJDD algorithm and FFT-QSPA with 50
iterations. It is seen that at a SER of
$10^{-6}$, the IJDD algorithm performs only 0.67dB away from the
FFT-QSPA. Similar observation can be made for WER.

To illustrate the rate of decoding convergence of the IJDD
algorithm, simulations were also carried out for $k_{max}=10$ and
$k_{max}=5$, respectively, with results shown in Fig.~\ref{Rate_conv}. It is
seen that with 10 and 50 iterations, the BER curves of the IJDD
algorithm nearly overlap each other. Even with 5 iterations, the
loss of performance is only 0.35 dB compared to IJDD with 50
iterations.

\begin{figure}
  \center
  \includegraphics[width=7cm]{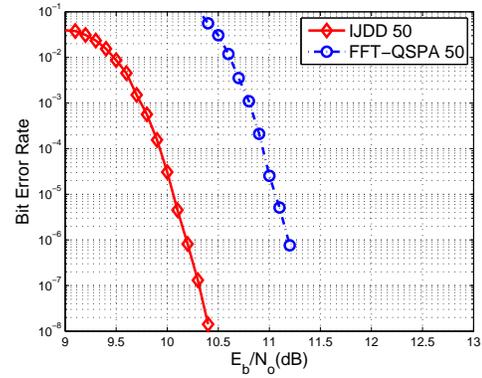}
  \caption{Error performance of the 32-ary (1023,781) LDPC code decoded with the FFT-QSPA and the IJDD algorithm (32-QAM)}\label{BER1023}
\end{figure}

\emph{Example 2}: Consider a 32-ary (1023, 781) regular LDPC code
constructed based on finite fields. The row redundancy is 781 and
both the row and column weights are 32. The performance of this
code incorporated with 32-QAM modulation over the AWGN channel is
illustrated in Fig.~\ref{BER1023}, where the codewords were obtained by
encoding the randomly generated source data. Surprisingly, the
IJDD algorithm outperforms the FFT-QSPA by 1.0 dB at a BER of $10^{-5}$. This
result may be caused by large column weights and large row
redundancy of the code, which can offer high reliability for IJDD.
In addition, the large column weights and large row redundancy of
the code may make the FFT-QSPA algorithm not suitable for
decoding it.

\section{Conclusions}
In this paper, we propose an iterative joint detection-decoding
algorithm for $q$-ary LDPC-coded systems, which can be
characterized as a hard-decision based message-passing algorithm
and so has low computational complexity. For $q$-ary LDPC codes
with large row redundancy and column weights, the proposed
algorithm can offer good performance or even outperforms the FFT-QSPA
with a lower computational complexity. Furthermore, the fast rate
of decoding convergence of our proposed algorithm makes it
particularly attractive, thus offering an attractive candidate for
practical applications of $q$-ary LDPC codes. Although only
regular LDPC codes are considered here, the proposed algorithm can
be extended directly to decode irregular $q$-ary LDPC codes.

\section*{Acknowledgment}
The authors would like to thank Prof. Shu Lin for his enlightening
lectures. They also wish to thank Chao Chen for providing the
parity-check matrices used in this paper, and Lin Zhou and Wei Lin
for helpful discussions. This work is supported jointly by NSFC
under Grants 60972046 and U0635003, the National S\&T Major
Special Project (No. 2009ZX03003-011), and the PCSIRT (No.
IRT0852).


\bibliographystyle{IEEEtran}

\end{document}